\documentclass[12pt,preprint]{aastex}

\newcommand{\etal}{{et al.~}}

\newcommand{\kpc}{\>{\rm kpc}}
\newcommand{\Mpc}{\>{\rm Mpc}}
\newcommand{\kms}{\>{\rm km}\,{\rm s}^{-1}}
\newcommand{\Msun}{\>{\rm M_{\odot}}}

\def\u{{\bf u}}
\def\v{{\bf v}}
\def\w{{\bf w}}
\def\x{{\bf x}}
\def\K{{\rm \ K}}

\def\gtsima{$\; \buildrel > \over \sim \;$}
\def\ltsima{$\; \buildrel < \over \sim \;$}
\def\gsim{\lower.7ex\hbox{\gtsima}}
\def\lsim{\lower.7ex\hbox{\ltsima}}
\def\simgt{\lower.7ex\hbox{\gtsima}}
\def\simlt{\lower.7ex\hbox{\ltsima}}
\def\la{\lsim}

\def\lta{\la}

\shortauthors{van den Bosch \etal}
\shorttitle{The Angular Momentum of Gas in Proto-Galaxies}

\received{2002 January 7}
\begin{document}

\title{The Angular Momentum of Gas in Proto-Galaxies: \\
I -- Implications for the Formation of Disk Galaxies}

\author{Frank C. van den Bosch}
\affil{Max-Planck Institut f\"ur Astrophysik, Karl
       Schwarzschild Str. 1, Postfach 1317, 85741 Garching, Germany}
\email{vdbosch@mpa-garching.mpg.de}

\author{Tom Abel\altaffilmark{1}, Rupert A. C. Croft\altaffilmark{2}, 
Lars Hernquist}
\affil{Harvard Smithsonian Center for Astrophysics, MA--02138,
       Cambridge, U.S.A.}

\and

\author{Simon D.M. White}
\affil{Max-Planck Institut f\"ur Astrophysik, Karl
       Schwarzschild Str. 1, Postfach 1317, 85741 Garching, Germany}

\altaffiltext{1}{Present address: Institute of Astronomy, Madingley 
                 Road, Cambridge, CB03 0HA, UK}
\altaffiltext{2}{Present address: Physics Department, Carnegie Mellon 
                 University, Pittsburgh, PA 15213, U.S.A.}

\begin{abstract}
  We use numerical  simulations of structure formation in  a Cold Dark
  Matter (CDM) cosmology to compare the angular momentum distributions
  of dark matter and non-radiative gas in a large sample of halos.  We
  show  that   the  two  components  have   identical  spin  parameter
  distributions and  that their angular  momentum distributions within
  individual halos  are very similar, all in  excellent agreement with
  standard assumptions.  Despite  these similarities, however, we find
  that the  angular momentum  vectors of the  gas and dark  matter are
  poorly  aligned,  with a  median  misalignment  angle  of $\sim  30$
  degrees,   which  might   have  important   implications   for  spin
  correlation  statistics used  in  weak lensing  studies. We  present
  distributions  for the  component of  the angular  momentum  that is
  aligned with the total angular  momentum of each halo, and find that
  for between 5 and 50 percent of the mass this component is negative.
  This  disagrees  with  the  generally  adopted  `Universal'  angular
  momentum  distribution, for  which the  mass fraction  with negative
  specific angular  momentum is zero.  We discuss  the implications of
  our  results for  the formation  of disk  galaxies.   Since galactic
  disks generally  do not contain counter-rotating stars  or gas, disk
  formation  cannot  occur  under  detailed conservation  of  specific
  angular  momentum.   We  suggest  that the  material  with  negative
  specific  angular momentum combines  with positive  angular momentum
  material  to  build a  bulge  component, and  show  that  in such  a
  scenario  the remaining  material can  form  a disk  with a  density
  distribution that is very close to exponential.
\end{abstract}


\section{Introduction}
\label{sec:intro}

Understanding  the  structure  and   formation  of  disk  galaxies  is
intimately linked to understanding the origin of its angular momentum.
In hierarchical  structure formation  scenarios the luminous  parts of
galaxies  form from  gas that  is cooling  and condensing  within dark
matter (DM) halos  and these merge to build  larger and larger objects
(White \& Rees  1978).  Within this picture, the  current paradigm for
disk formation  contains three important ingredients:  (i) the angular
momentum originates  from cosmological torques (Hoyle  1953), (ii) the
gas  and dark matter  within virialized  systems have  initial angular
momentum distributions  that are identical (Fall  \& Efstathiou 1980),
and (iii) the gas conserves its specific angular momentum when cooling
(Mestel 1963).

It  is  well  established  that cosmological  torques  impart  angular
momentum  to dark matter  halos.  Many  studies have  investigated the
resulting  angular momenta  of dark  matter halos  using  either tidal
torque  theory (e.g.,  Peebles  1969; Doroshkevich  1970; White  1984;
Catelan  \&  Theuns 1996)  or  N--body  simulations  (e.g., Barnes  \&
Efstathiou 1987; Warren \etal  1992; Bullock \etal 2001). The specific
angular  momentum   of  halos   is  typically  parameterized   by  the
dimensionless spin parameter
\begin{equation}
\label{lam}
\lambda \equiv \frac{|{\bf J}| \, |E|^{1/2}}{G \, M^{5/2}},
\end{equation}
with $G$  the gravitational constant and  ${\bf J}$, $E$,  and $M$ the
total angular  momentum, energy and  mass of the  object, respectively
(Peebles 1969).  Under the assumptions (ii) and (iii) disks have scale
lengths $R_d \propto \lambda R_{\rm vir}$.  Given the distributions of
halo  spin parameters  and of  halo  virial radii  $R_{\rm vir}$,  the
implied distribution  of disk scale lengths is  in excellent agreement
with observations  (e.g., Dalcanton, Spergel \& Summers  1997; Mo, Mao
\& White 1998; de Jong \& Lacey 2000).  This success has prompted many
detailed  studies of  disk galaxy  formation, always  under  the three
assumptions  listed  above (van  den  Bosch  1998,  2000, 2001,  2002;
Jimenez \etal  1998; Natarajan 1999; Heavens \&  Jimenez 1999; Firmani
\& Avila-Reese  2000; Avila-Reese \& Firmani  2000; Buchalter, Jimenez
\& Kamionkowski 2001).

The standard  picture of  disk formation that  has emerged  from these
studies has been remarkably successful in explaining a wide variety of
observational properties  of disk galaxies.   However, two significant
problems, both  related to the  angular momentum, have come  to light.
First of  all, detailed hydro-dynamical simulations  of disk formation
in a cold dark matter (CDM)  Universe yield disks that are an order of
magnitude  too small  (Navarro \&  Benz 1991;  Navarro \&  White 1994;
Steinmetz \& Navarro 1999;  Navarro \& Steinmetz 2000).  This problem,
known as  the angular  momentum catastrophe, is  a consequence  of the
hierarchical formation of galaxies which  causes the baryons to lose a
large fraction  of their angular momentum  to the dark  matter.  It is
unclear whether  this is a  problem for the  theory or merely  for the
particular  way  in  which  feedback  processes  are  implemented  (or
ignored) in the simulations (e.g., Navarro \& White 1994; Weil, Eke \&
Efstathiou  1998;  Sommer-Larsen, Gelato  \&  Vedel  1999; Thacker  \&
Couchman 2001).

The  second problem concerns  the actual  density structure  of galaxy
disks.    Assuming  detailed   angular  momentum   conservation,  this
structure  is a  direct  reflection of  the  distribution of  specific
angular momentum in the proto-galaxy.  Bullock \etal (2001) determined
the angular  momentum distributions  of individual dark  matter halos,
which  according to assumption  (ii) should  also reflect  the angular
momentum distribution of the gas  that forms the disk.  However, these
distributions seem to have far  too much low angular momentum material
to be consistent with the typical exponential density distributions of
disk galaxies (Bullock \etal 2001;  van den Bosch 2001; van den Bosch,
Burkert \& Swaters 2001). Knebe  \etal (2002) and Bullock, Kravtsov \&
Col\'{\i}n (2002)  investigated the angular momentum  profiles of warm
dark  matter (WDM)  halos.   Except  for a  slight  difference in  the
average spin  parameter, CDM and  WDM halos have very  similar angular
momentum distributions,  implying that this  particular problem cannot
be solved  by invoking WDM instead  of CDM. A  more promising solution
has  recently been  suggested  by  Maller \&  Dekel  (2002). In  their
picture  halo angular  momentum  originates from  the orbital  angular
momentum of  merging satellites, rather  than from pure  tidal torques
(see also  Maller, Dekel \&  Somerville 2001). Combining  this picture
with a simple  model for feedback and tidal  stripping Maller \& Dekel
are able  to explain the  angular momentum profiles of  dwarf galaxies
measured by van den Bosch \etal (2001).

These angular  momentum problems seem  to imply that  assumptions (ii)
and (iii) listed  above, and which have been  essential ingredients of
our  standard paradigm  for the  formation  of disk  galaxies, may  be
incorrect.  Remarkably, despite many papers that have investigated the
angular momentum  distributions within dark matter halos,  to the best
of our knowledge, no study  so far has addressed the spin distribution
of the  gas in these halos in  order to test whether  the {\it ansatz}
that the  gas and  dark matter have  similar initial  distributions of
specific  angular momentum is  correct.  In  fact, it  seems plausible
that they  may not.  Gas  and dark matter suffer  different relaxation
mechanisms  during halo  collapse: Whereas  the dark  matter undergoes
collisionless  virialization  through   violent  relaxation,  the  gas
attempts  to achieve  a  hydrostatic equilibrium  through shocks.   In
order to  test whether the  initial angular momentum  distributions of
gas  and  dark  matter   are  indeed  similar,  we  perform  numerical
simulations  of  structure   formation  in  a  $\Lambda$CDM  cosmology
including both dark matter  and a non-radiative gas.  After presenting
our    simulations     and    outlining    our     analysis    methods
(Section~\ref{sec:method}),  we  present  various  statistics  of  the
angular   momentum   distributions  (Section~\ref{sec:results}).    In
Section~\ref{sec:diskform} we discuss  the implications of our results
for the theory of disk  galaxy formation, and we summarize our results
in Section~\ref{sec:concl}.

\section{Method}
\label{sec:method}

\subsection{Numerical Simulations}
\label{sec:simul}

We use the public version of the smoothed particle hydrodynamics (SPH)
code  GADGET  developed  by  V.   Springel  (1999)  and  described  in
Springel,  Yoshida \&  White  (2000).  The  chosen  flat cosmology  is
vacuum energy  dominated with $\Omega_\Lambda=0.7$  and matter content
$\Omega_m=0.3$ with a baryon density $\Omega_{b} = 0.021 h^{-2}$ and a
Hubble constant of $67 \kms  \Mpc^{-1}$ (i.e., $h=0.67$).  We choose a
relatively small  box of $10 h^{-1}$  comoving Mpc and  evolve it from
redshift 59  down to redshift  three.  We use  identical power-spectra
for  baryons  and  dark  matter   given  by  the  fitting  formula  of
Efstathiou, Bond  \& White (1992)  and use $128^3$ particles  each for
the  gas  and  the  dark  matter component.   The  power  spectrum  is
normalized to have  a typical mean overdensity of  $\sigma_8 = 0.9$ at
redshift zero  in a top hat sphere  of $8 h^{-1} \Mpc$.   Gas and dark
matter  particles have  masses of  $m_{\rm gas}  = 6.26  \times 10^{6}
h^{-1}  \Msun$ and  $m_{\rm DM}  = 3.34  \times 10^{7}  h^{-1} \Msun$,
respectively.  The  gravitational softening  lengths for both  the gas
and the dark matter were chosen  to be $4 h^{-1} \kpc$ (comoving). The
gas has an initial entropy corresponding to a temperature of $2 \times
10^4  \K$  for  a  mean  molecular  weight of  one  proton  mass.   An
artificial  temperature  floor at  $10^4\K$  (for  the same  molecular
weight) is introduced to mimic the heating effects from re-ionization.
Both this temperature floor and the initial temperature are too low to
play a  significant role  in the simulated  dynamics because  they are
significantly  lower than  the  virial temperatures  of  the halos  we
analyze. This  is evident in our  analysis as the  reported results do
not depend  on the mass of  the halos studied.   All results presented
below correspond to the final output at $z=3$.

The robustness  of our results with respect  to resolution, simulation
box   size,   and  the   redshift   of   analysis   is  addressed   in
Section~\ref{sec:robust}.

\subsection{The Halo Identification Algorithm}
\label{sec:halos}

We start by using the  publicly available group finder HOP, written by
Eisenstein \&  Hut (1998), to  identify overdensity peaks in  the dark
matter distribution  using a  threshold density of  20 times  the mean
dark matter density.  In each  group we identify the densest particle,
and  we determine  the virial  radius $R_{\rm  vir}$ of  the spherical
volume, centered on this densest particle, inside of which the average
dark  matter  density  is  $\Delta_{\rm vir}(z)$  times  the  critical
density $\rho_{\rm  crit}(z)$.  For our  adopted cosmology and  at the
redshift of the  output analyzed $\Delta_{\rm vir} =  174.7$ (Bryan \&
Norman 1998).  Next  we compute the center-of-mass of  all dark matter
particles inside this spherical volume which we associate with the new
center of the  halo. We compute the new $R_{\rm  vir}$ around this new
halo  center, as  well  as the  associated  new center-of-mass.   This
procedure is  iterated until  the distance between  the center-of-mass
and the  adopted halo center  is less than  one percent of  the virial
radius $R_{\rm vir}$.  Typically this requires  of the order of 2 to 5
iteration steps.

All  dark matter  and  gas particles  inside  the resulting  spherical
volume  with  radius  $R_{\rm   vir}$  are  considered  halo  members.
Although our  iterative method assures that the  center-of-mass of the
dark matter component is similar  to the adopted halo center, the same
is not necessarily true for the gas component. We therefore remove all
halos from our  sample for which the distance  between the halo center
and  the center-of-mass  of  the  gas particles  is  larger than  $10$
percent  of  $R_{\rm vir}$.   We  checked  that  none of  our  results
correlate with the  offset between the centers of mass  of the gas and
dark matter. In addition, we repeated the entire analysis adopting the
position  of the most  bound particle  as halo  center and  found very
simular results.

In order  to allow sufficiently  accurate measurements of  the angular
momentum vectors of the gas and  dark matter we only accept halos that
have  more than $100$  gas particles  {\it and}  more than  $100$ dark
matter particles.   In addition, if  any two halos overlap,  i.e., the
distance between the halo centers is less than the sum of their virial
radii, we remove the least  massive halo from our sample.  This leaves
a total of  378 halos of which  259 are isolated in that  they have no
subgroups detected, and  they do not overlap with  any other halo.  Of
these 378 halos  23 have more than 5000 particles  (in total), and are
used  for a  more detailed  analysis of  the distribution  of specific
angular momentum (see Section~\ref{sec:distribution}).

\subsection{Theoretical Background}
\label{sec:theory}

Our main goal is to  compare the angular momentum distributions of the
gas and  dark matter components. However,  one cannot do  so by simply
computing angular  momentum distributions  from the velocities  of the
gas and  dark matter particles. Both  the dark matter and  the gas are
fluids, for which the velocity  of each {\it microscopic} particle can
be written as $\v = \u +  \w$.  Here $\u$ is the mean streaming motion
at  the location $\x$  of the  microscopic particle,  and $\w$  is the
particle's random motion.   The macroscopic, collisionless dark matter
particles  in $N$-body  simulations can  be considered  a  Monte Carlo
realization   of  the   phase-space  distribution   function   of  all
microscopic  particles.   The velocities  of  these  particles in  the
simulation  correspond to $\v$  and are  computed by  solving Newton's
equations of  motion.  In the case  of the gas,  however, the Smoothed
Particle  Hydrodynamics  (SPH)  approach  is used,  which  yields  the
streaming  motions $\u$  of the  gas  particles by  solving the  Euler
equations (with some artificial  viscosity to take account of shocks).
Information about the  random motion of the gas  particles is provided
by the internal energy of each particle.

Thus the velocities of the  dark matter particles and gas particles in
the  simulation correspond to  different motions.  This must  be taken
into  account before comparing  their angular  momentum distributions.
This is most  easily achieved by computing the  velocities $\v$ of the
gas  particles  from their  streaming  motions  and their  (isotropic)
velocity dispersion  tensor.  For each  gas particle we  randomly draw
$100$ random velocities $\w$ which  we add to the particle's streaming
motion $\u$.   For each of the  three Cartesian components  of $\w$ we
draw a velocity from a Gaussian with a standard deviation given by
\begin{equation}
\label{veldisp}
\sigma = \sqrt{k \, T \over \mu} = \sqrt{2 \, U \over 3}
\end{equation}
Here $U$ and $T$ are the internal energy per unit mass and temperature
of the  gas particle, respectively,  $k$ is Boltzmann's  constant, and
$\mu$ is  the mean  molecular weight of  the gas. After  this `thermal
broadening'  we  find that  the  rms velocities  $\langle  |  \v |  ^2
\rangle$  of the  gas  and dark  matter  particles are,  for the  vast
majority of our halos, very similar, indicating that the halos are, to
a good approximation, in hydrostatic equilibrium.

We  thus use  the  velocities  $\v$ to  compare  the angular  momentum
distributions of  the gas and dark matter.   However, the distribution
that  is of  interest for  galaxy  formation is  the angular  momentum
distribution based on the streaming motions $\u(\x)$.  After all, when
the gas  cools $\v \rightarrow  \u$, i.e., the typical  random motions
become negligible, at least  in sufficiently massive halos.  Under the
assumption that  the gas conserves its specific  angular momentum, the
resulting disk will thus have an angular momentum distribution related
to its  streaming motions  $\u$. Ideally one  would therefore  like to
compare the angular momentum distributions computed from the streaming
motions  $\u(\x)$.  However,  this requires  estimating  the streaming
motions of  the dark  matter. In principle  this could be  achieved by
smoothing  the dark  matter velocities  $\v$, but  it is  unclear what
smoothing scale  to use  and resolution issues  are likely to  play an
important  role.  We  therefore do  not  attempt to  compute the  dark
matter streaming motions,  but instead use the velocities  $\v$ of the
dark matter  and gas particles  when comparing their  angular momentum
distributions.  However,  the streaming motions  of the gas  {\it are}
available,  and  we  present  a  detailed comparison  of  the  angular
momentum  distributions  of  the  gas   based  on  $\v$  and  $\u$  in
Section~\ref{sec:distribution}.

\subsection{Analysis}
\label{sec:angmom}

For each halo we compute the  total angular momenta of the gas and the
dark matter particles, using,
\begin{equation}
\label{angmom}
{\bf  J}_{\rm gas,  DM} =  m_{\rm gas,DM}  \sum_{i=1}^{N_{\rm gas,DM}}
        {\bf r}_i \times ({\bf v}_i - {\bf \bar{v}}_{\rm com}).
\end{equation}
Here ${\bf  \bar{v}_{\rm  com}}$ is the mean  (bulk)  velocity of  the
entire halo (dark   matter plus gas) and ${\bf   r}_i$  is the  radial
vector with  respect  to the halo   center.  We  follow Bullock  \etal
(2001) and define the following modified spin parameters
\begin{equation}
\label{lambda}
\lambda'_{\rm gas, DM} = \frac{|{\bf j}_{\rm gas,DM}|}{\sqrt{2} \, 
R_{\rm vir} \, V_{\rm vir}}.
\end{equation}
Here  ${\bf j}_{\rm  gas}$ and  ${\bf  j}_{\rm DM}$  are the  specific
angular momenta of the gas  and dark matter, respectively, and $V_{\rm
vir}  =  \sqrt{G (M_{\rm  gas}  +  M_{\rm  DM})/R_{\rm vir}}$  is  the
circular velocity at the virial radius $R_{\rm vir}$.  This definition
of the spin parameter has the advantage that it does not depend on the
energy  content  of  the  halo  which can  be  difficult  to  estimate
reliably.   In  what  follows  we  always  adopt  the  spin  parameter
definition of equation~(\ref{lambda}), and we drop the prime.

In addition to the  spin parameters of the dark  matter and the gas we
also compute the angle
\begin{equation}
\theta =  \cos^{-1} \left[ \frac{{\bf J}_{\rm gas}  \cdot {\bf J}_{\rm
DM}} {|{\bf J}_{\rm gas}|\, |{\bf J}_{\rm DM}|} \right]
\end{equation}
between  the angular momentum  vectors of  the two  components.  After
rotating the coordinate  frame such that the $z$-axis  is aligned with
the angular momentum  vector of either the dark matter  or the gas, we
also  compute the  fractions of  dark  matter and  gas particles  with
negative  angular  momentum  about  their  respective  $z$-directions.
These  fractions  are  denoted  by  $f_{\rm DM}$  and  $f_{\rm  gas}$,
respectively,  and give  an  indication of  the  amount of  disordered
motion relative to the ordered motion.

Note that since at each position  the expectation value of $\w$ is, by
definition,  the  null  vector,  the values  of  $\lambda_{\rm  gas}$,
$\lambda_{\rm  DM}$  and  $\theta$  are  independent  of  whether  one
considers  the streaming  motions $\u$  or the  microscopic velocities
$\v$.  However, the angular  momentum {\it distributions} may be quite
different.  In order  to  distinguish  these two  cases  we label  the
appropriate   quantities  with  the   superscript  $v$   whenever  the
microscopic  velocities  $\v$ are  used.  Thus  a particle's  specific
angular momentum is  given by either ${\bf j} = {\bf  r} \times \u$ or
${\bf j}^{v} = {\bf r} \times \v$.

\section{Results}
\label{sec:results}

\begin{figure}
\plotone{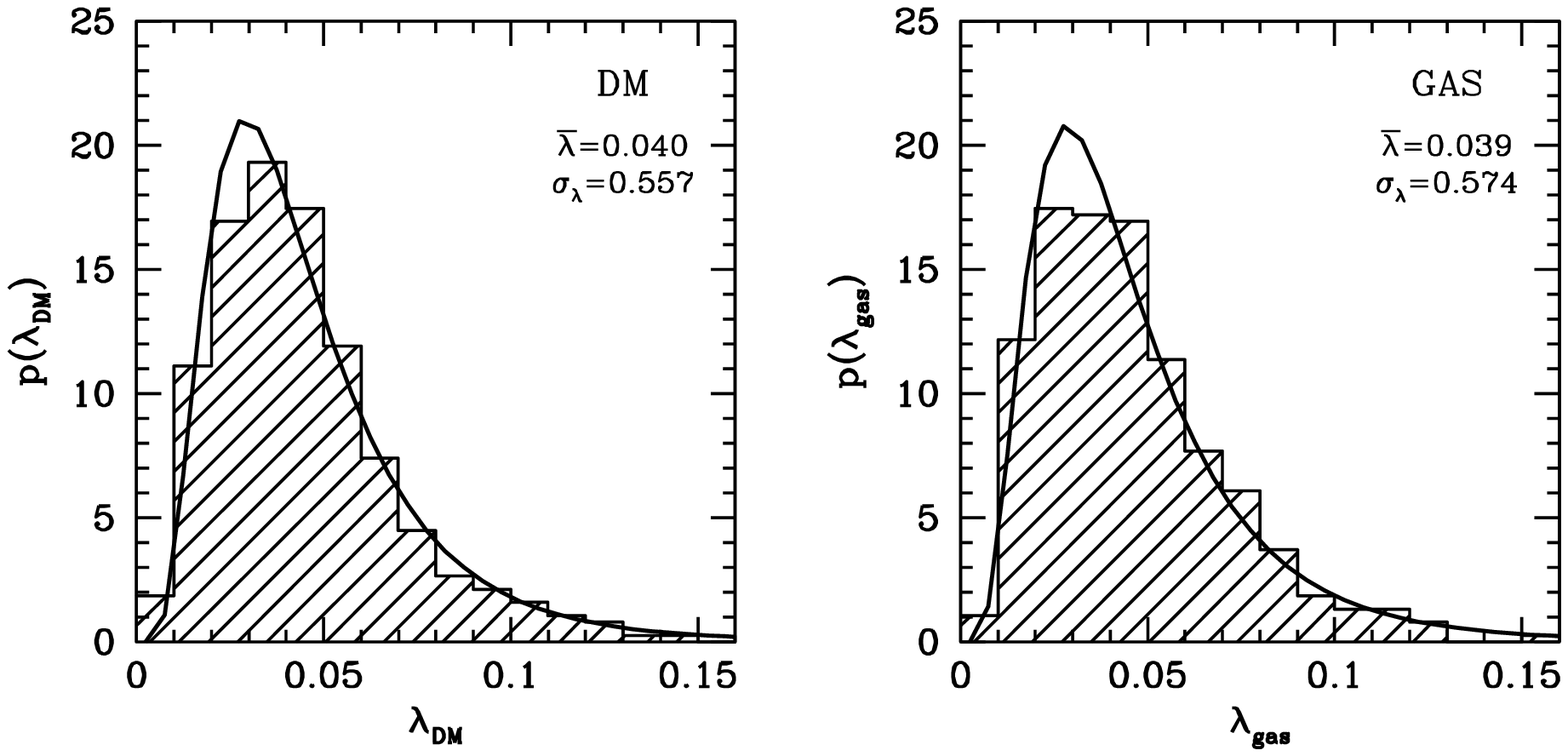}
\caption{The  distributions of  $\lambda_{\rm DM}$  (right  panel) and
$\lambda_{\rm  gas}$ (left  panel) for  all 378  halos in  our sample.
Solid   lines   are  the   best   fit   log-normal  distributions   of
equation~(\ref{spindistr}). The values of  the best fit parameters are
indicated in each panel.  Note that  the gas and dark matter have very
similar distributions of spin parameters.}
\label{fig:histo} 
\end{figure}

Figure~\ref{fig:histo} plots the  histograms of $\lambda_{\rm DM}$ and
$\lambda_{\rm  gas}$   for  all  378   halos  in  our   sample.   Both
distributions are well fit by a log-normal distribution
\begin{equation}
\label{spindistr}
p(\lambda){\rm d} \lambda = {1 \over \sqrt{2 \pi} \sigma_{\lambda}}
\exp\biggl(- {{\rm ln}^2(\lambda/\bar{\lambda}) \over 2
  \sigma^2_{\lambda}}\biggr) {{\rm d} \lambda \over \lambda}.
\end{equation}
For  the dark  matter we  find  $\bar{\lambda}_{\rm DM}  = 0.040$  and
$\sigma_{\lambda_{\rm    DM}}   =   0.56$,    while   for    the   gas
$\bar{\lambda}_{\rm  gas} = 0.039$  and $\sigma_{\lambda_{\rm  gas}} =
0.57$.   Clearly,   the  distributions  of   $\lambda_{\rm  gas}$  and
$\lambda_{\rm DM}$ are indistinguishable.  If we only consider the 259
isolated halos we obtain very  similar distributions both for the dark
matter and for the gas.

\begin{figure}
\epsscale{.8}
\plotone{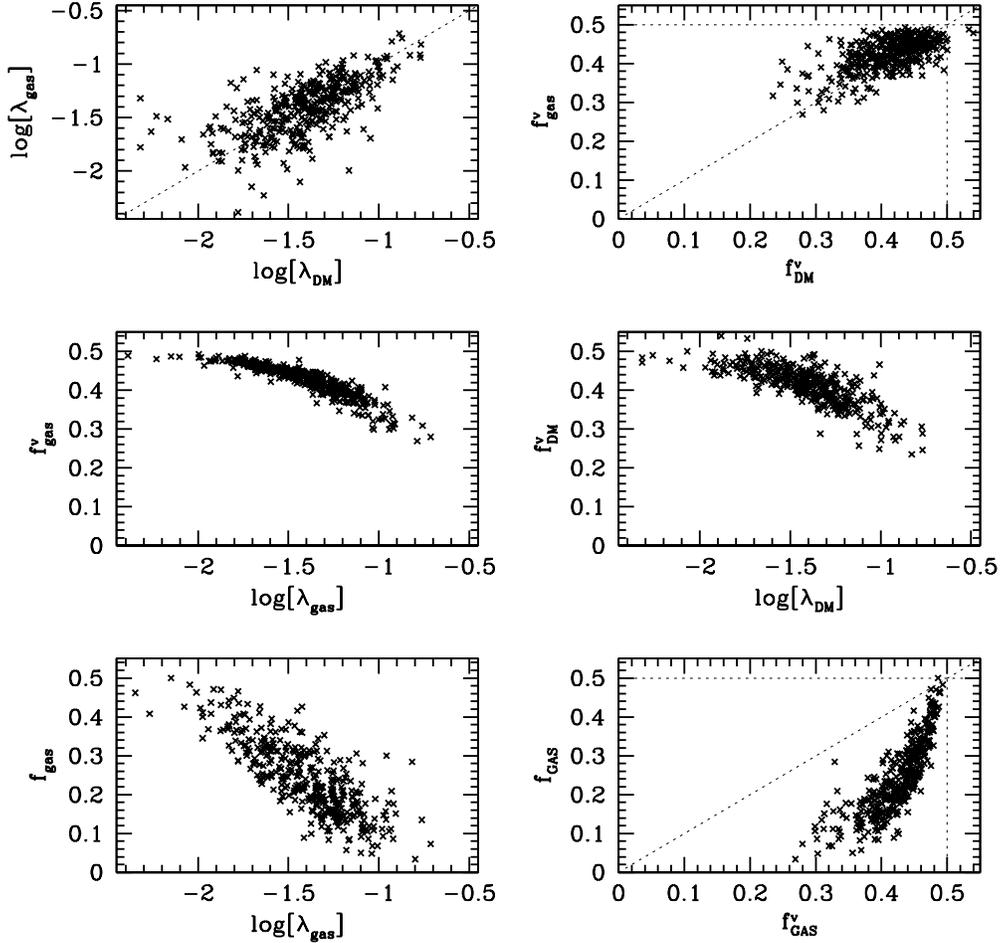}
\caption{A comparison of various global angular momentum parameters of
the dark  matter and gas for all  378 halos in our  sample.  The upper
left panel  plots $\lambda_{\rm  gas}$ versus $\lambda_{\rm  DM}$. The
dotted line corresponds to perfect agreement (i.e., $\lambda_{\rm gas}
=  \lambda_{\rm  DM}$) and  is  plotted  for  comparison. Although  on
average the  spin parameters of gas  and dark matter  are similar, for
each  individual halo the  difference between  the spin  parameters of
both components can  be relatively large. In the  upper right panel we
plot $f^v_{\rm gas}$ versus $f^v_{\rm DM}$. Again, on average the mass
fractions with negative specific  angular momentum are similar for the
gas  and the dark  matter (i.e.,  $\langle f^v_{\rm  DM}/f^v_{\rm gas}
\rangle = 0.97 \pm 0.10$). The two panels in the middle plot $f^v_{\rm
gas}$  versus $\lambda_{\rm  gas}$  (left) and  $f^v_{\rm DM}$  versus
$\lambda_{\rm  DM}$  (right).  Note  how  the  total specific  angular
momentum of each component is anti-correlated with its respective mass
fraction with negative $j_z$.  The  same is true when plotting $f_{\rm
gas}$  rather than $f^v_{\rm  gas}$ (lower  left panel).  Finally, the
lower  right panel  plots  $f_{\rm gas}$  versus  $f^v_{\rm gas}$.  As
expected, all halos  have $f_{\rm gas} < f^v_{\rm  gas}$. Note however
that  the   difference  depends  on   $f^v_{\rm  gas}$  and   thus  on
$\lambda_{\rm gas}$. See the text for a more detailed discussion.}
\label{fig:lambda}
\end{figure}

In   the  upper   left  panel   of  Figure~\ref{fig:lambda}   we  plot
$\lambda_{\rm gas}$ versus $\lambda_{\rm DM}$.  Although $\lambda_{\rm
gas}$ and $\lambda_{\rm DM}$ are well correlated, with a Spearman rank
coefficient $r_s = 0.67$, which has a probability of virtually zero to
occur  under the  null hypothesis  of no  correlation, the  scatter is
relatively large.   The average  of $\lambda_{\rm gas}  - \lambda_{\rm
DM}$ is $-0.001$ with a  standard deviation of $0.020$, which confirms
once again  that on average  the spin parameters  of the gas  and dark
matter  are very  similar.  However,  on  a one-to-one  basis the  two
components can  have quite  different specific angular  momenta; i.e.,
the average  (median) of $|\lambda_{\rm  gas} - \lambda_{\rm  DM}|$ is
$0.015$ ($0.011$). We find  no significant trend between $\lambda_{\rm
gas} -  \lambda_{\rm DM}$ and $M_{\rm  vir}$ (see upper  left panel of
Figure~\ref{fig:theta_corr}) indicating that  the scatter is real, and
not an artifact of discreteness noise.

The  upper right  panel of  Figure~\ref{fig:lambda}  plots $f^{v}_{\rm
gas}$  versus $f^{v}_{\rm DM}$.   With $\langle  f^v_{\rm DM}/f^v_{\rm
gas}  \rangle =  0.97  \pm 0.10$  the  two mass  components have  very
similar mass  fractions with negative specific  angular momentum.  The
middle  panels  of Figure~\ref{fig:lambda}  show  that  the values  of
$f^v_{\rm gas}$  and $f^v_{\rm DM}$ are  strongly anti-correlated with
their respective spin parameters.   A similar anti-correlation is also
present for the  gas when $f_{\rm gas}$, is  used instead of $f^v_{\rm
gas}$ (lower left  panel).  Again, none of these  results change if we
only focus on the 259 isolated halos. These anti-correlations indicate
that  it  is the  {\it  amount} of  ordered  motion  that is  directly
responsible for the amplitude of the total specific angular momentum.

Finally,  the  lower  right  panel  of  Figure~\ref{fig:lambda}  plots
$f_{\rm gas}$  versus $f^v_{\rm gas}$.   For all halos $f_{\rm  gas} <
f^v_{\rm  gas}$, illustrating  the trivial  result that  the streaming
motions of  the gas are  more ordered than their  microscopic motions.
Note,  however, that  there is  not simply  a constant  offset between
$f_{\rm  gas}$ and  $f^v_{\rm gas}$.   Rather, the  difference between
$f_{\rm  gas}$ and  $f^v_{\rm gas}$  is larger  for halos  with larger
$\lambda_{\rm  gas}$, and the  distribution of  $f_{\rm gas}$  is much
broader  than  that  of   $f^v_{\rm  gas}$.  All  these  results  have
implications for the formation of disk galaxies, as we discuss in more
detail in Section~\ref{sec:diskform}.

\subsection{Misalignment of Angular Momentum Vectors}
\label{sec:misalign}

\begin{figure}
\epsscale{1.0}
\plotone{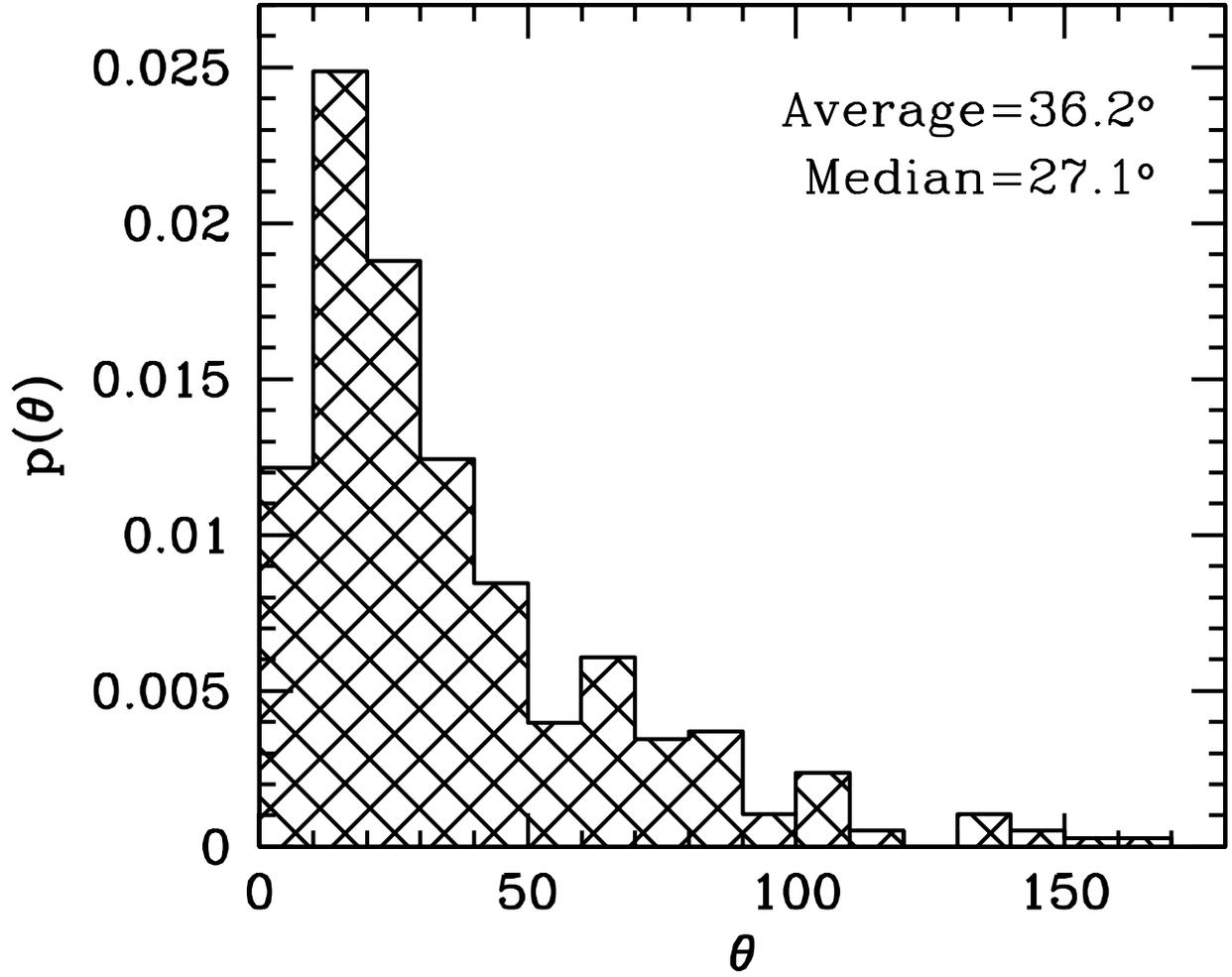}
\caption{The distribution  of the angle $\theta$  between the total
angular momentum vectors  of the dark matter and  the gas. The average
and mean of the distribution are indicated.}
\label{fig:theta}
\end{figure}

Figure~\ref{fig:theta}  plots  the  distribution of  the  misalignment
angle $\theta$ (in degrees) between the total angular momentum vectors
of the  dark matter  and the  gas.  The mean  ($\sim 36^{\rm  o}$) and
median ($\sim  27^{\rm o}$)  of the distribution  are similar  for the
entire sample and for the subsample of isolated halos.

\begin{figure}
\plotone{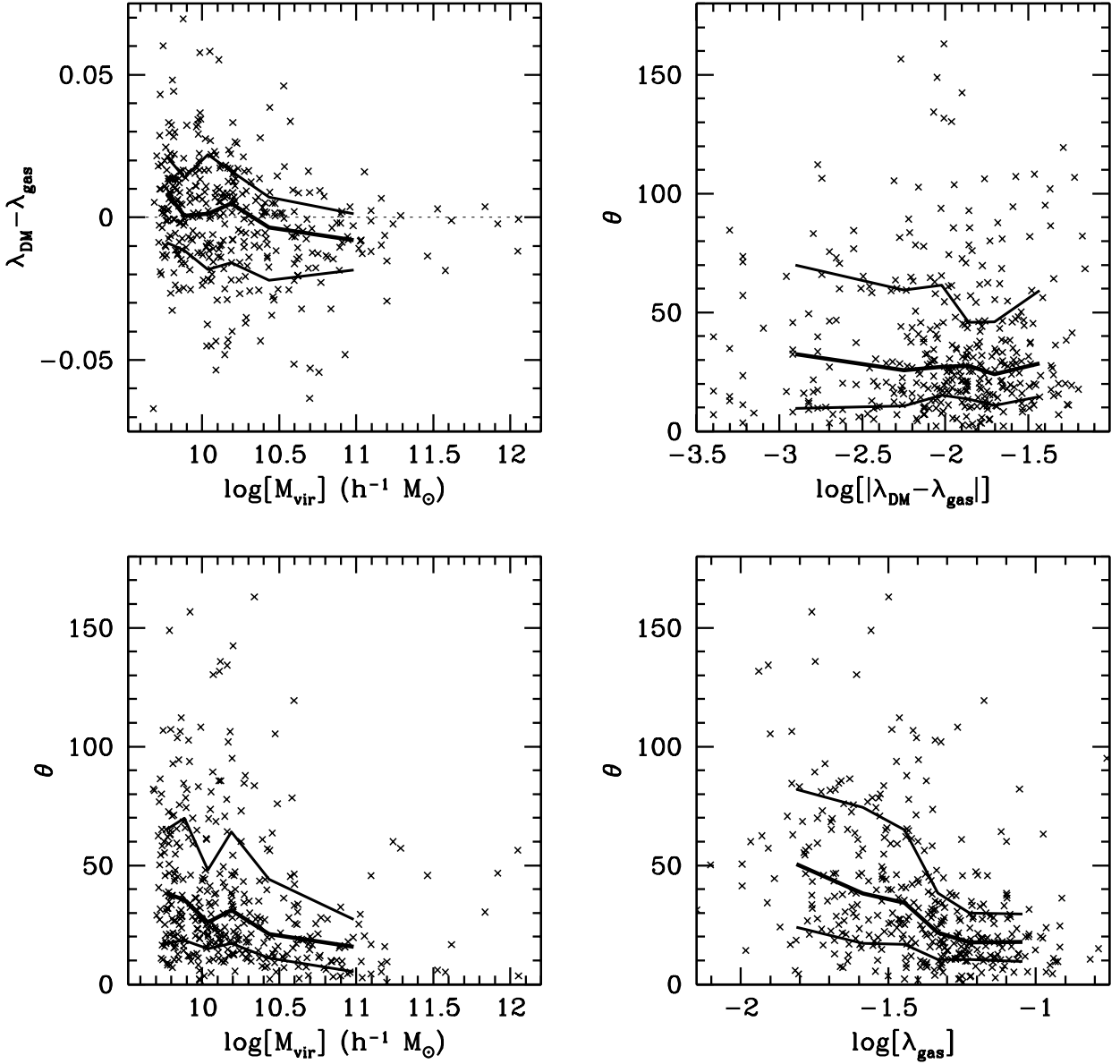}
\caption{The upper  left panel plots $\lambda_{\rm  DM} - \lambda_{\rm
gas}$ as function of the halo virial mass. The thick solid line in the
middle  indicates   the  run  of  the  average   $\lambda_{\rm  DM}  -
\lambda_{\rm gas}$, while the upper and lower solid lines indicate the
68 percent interval of $\lambda_{\rm DM} - \lambda_{\rm gas}$ at fixed
$M_{\rm vir}$. Note that the  difference in the spin parameters of the
gas and the dark matter do not depend significantly on the halo virial
mass.   The other  panels plot  $\theta$ as  function of  the absolute
difference between the  spin parameters of gas and  dark matter (upper
right panel), the  total virial mass (lower left  panel), and the spin
parameter  of  the  gas   (lower  right  panel).   Thick  solid  lines
correspond to  the mean and  68 percent interval of  $\theta$. Whereas
there  is   no  dependence  of   $\theta$  on  $|\lambda_{\rm   DM}  -
\lambda_{\rm  gas}|$,  the average  misalignment  between the  angular
momentum vectors of the gas and  dark matter is larger in less massive
halos and in halos with smaller spin parameters.}
\label{fig:theta_corr}
\end{figure}

Contrary to what one might expect, there is no correlation between the
misalignment  angle  $\theta$  and  the  absolute  difference  between
$\lambda_{\rm gas}$ and $\lambda_{\rm  DM}$.  This is evident from the
upper  right panel  of Figure~\ref{fig:theta_corr}.   The  solid lines
indicate the  mean and  the 68 percent  interval of $\theta$.   We do,
however,  find  a small  trend  that  the  average misalignment  angle
increases  with decreasing  halo mass  and spin  parameter  (see lower
panels of  Figure~\ref{fig:theta_corr}).  We  stress that there  is no
correlation  between $M_{\rm  vir}$  and $\lambda_{\rm  gas}$ so  that
these are two independent trends.  Both are likely due to discreteness
effects, in that the direction  of the angular momentum vectors can be
more accurately determined if either the total angular momentum or the
number of  halo particles is  larger.  However, since both  trends are
weak  and  the  average  misalignment  angle  is  still  significantly
different from zero  even for the most massive halos  and for the ones
with the  largest spin  parameters, we conclude  that there is  a true
misalignment  between the  angular momenta  of  the gas  and the  dark
matter (e.g., see column~(9) of Table~1).

This  misalignment may  have  important implications  for galaxy  spin
correlation  statistics.   In  the  past,  both  tidal  torque  theory
(Crittenden \etal  2001; Heavens, Refegier \&  Heymans 2000; Natarajan
\etal 2001) and pure N--body  simulations (Croft \& Metzler 2000) have
been  used to  calculate the  possibility of  intrinsic  alignments of
galaxy shapes, which may be  responsible for part of the measured weak
lensing signal (e.g.  van Waerbecke \etal 2000; McKay \etal 2001), and
may be  used to reconstruct the  shear field (Lee \&  Pen 2000, 2001).
In  these studies  the assumption  is made  that the  angular momentum
vectors  of the galaxies  are aligned  with those  of the  dark matter
halos.  The relatively large  misalignment angles found here, however,
may  diminish any  alignment of  galaxy shapes  that one  might expect
based on tidal torque theory.  We emphasize, however, that simulations
of higher  resolution are  required to reduce  the possible  impact of
discreteness effects on the distribution of misalignment angles.

\subsection{The Distribution of Angular Momentum}
\label{sec:distribution}

Although $\sim 100$ particles is  enough to obtain a reliable estimate
of the spin  parameter, many more particles are  required to determine
angular  momentum  distributions  (hereafter AMDs)  within  individual
haloes.  We  therefore limit  ourselves here to  halos with  more than
5000 particles in total, of which there are 23 in our sample.

We define the normalized specific angular momentum $l \equiv j/(R_{\rm
vir} V_{\rm vir})$ and  compute separate distributions $p(l)$ for dark
matter particles and  for gas particles. Here and  in what follows $j$
refers to the specific angular  momentum along the $z$-axis. Note that
$R_{\rm  vir} V_{\rm vir}$  is the  maximum specific  angular momentum
that a  halo particle can have  (i.e., $l \in  [-1,1]$).  We normalize
$p(l)$ to unity so that
\begin{equation}
\label{spinpl}
\lambda = {1 \over \sqrt{2}} \int_{-1}^{1} p(l) \, l \, {\rm d}l
\end{equation}

\begin{figure}
\plotone{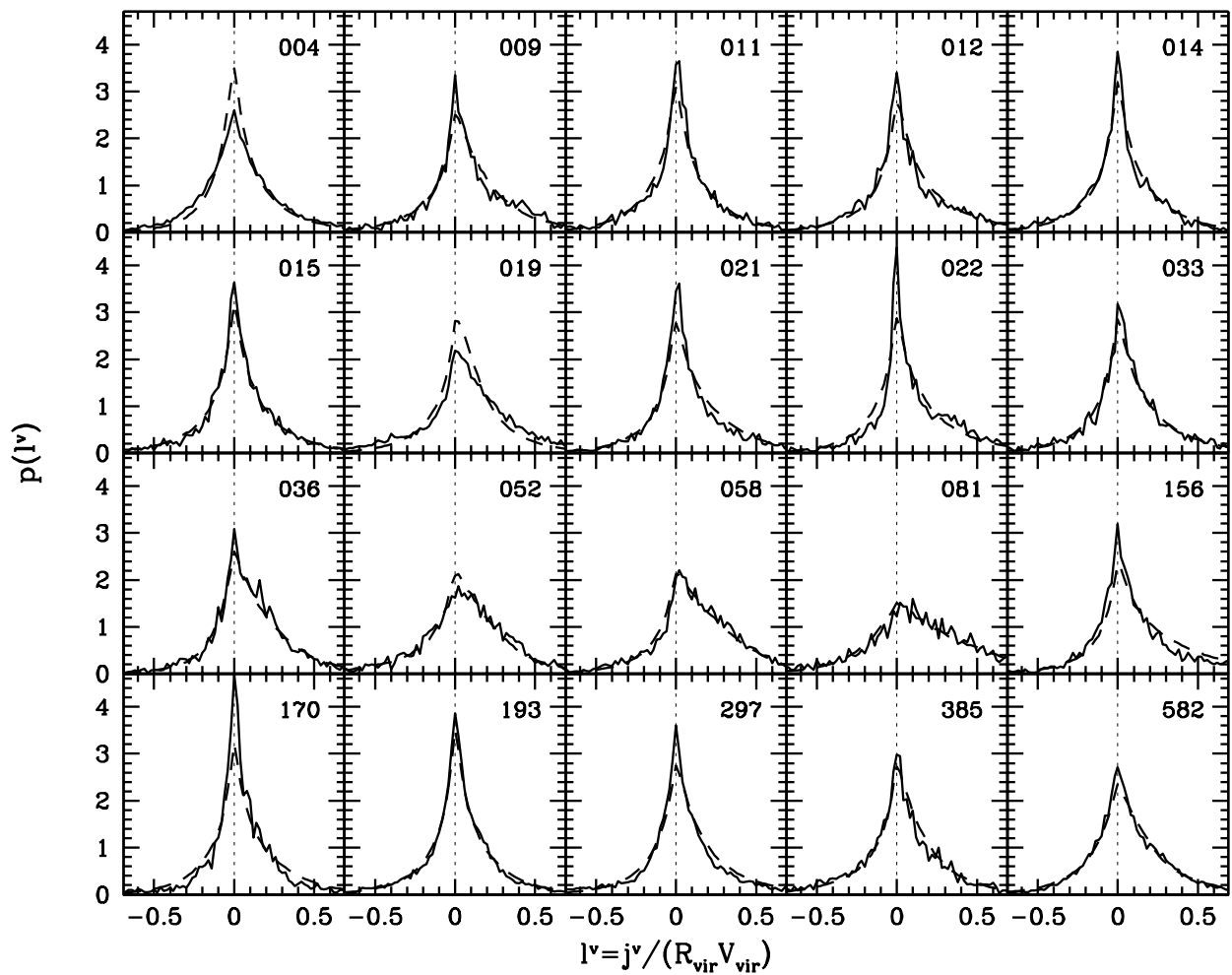}
\caption{The angular momentum distributions  $p(l^v)$ for the 20 halos
listed  in Table~1.   Solid and  dashed lines  correspond to  the dark
matter and  gas, respectively. Note the  remarkable similarity between
the angular momentum distributions of the gas and the dark matter.}
\label{fig:groups}
\end{figure}

In Figure~\ref{fig:groups}  we plot $p(l^v)$ both for  the gas (dashed
lines) and  for the dark matter  (solid lines) of $20$  halos that are
selected randomly from the sample  of $23$.  Some global parameters of
these $20$ halos  are listed in Table~1.  As can be  seen, the AMDs of
the two  mass components  are remarkably similar.   The fact  that the
$p(l^v)$ of  the gas appear less noisy  than for the dark  matter is a
consequence of  the fact that  because of the `thermal  broadening' of
the gas velocity  field we effectively have on  average 100 times more
gas   particles   per   halo   than   dark   matter   particles   (see
Section~\ref{sec:theory}).  In  a few  cases small differences  can be
discerned, but  there is no  indication for any  systematic difference
between  the AMDs  of  the two  components.   Therefore, the  standard
assumption that the gas and dark matter have the same distributions of
specific angular  momentum is to  first order correct, and  the shocks
that  occur  during the  virialization  process  do not  significantly
modify the angular momentum distribution of the gas.

\begin{figure}
\plotone{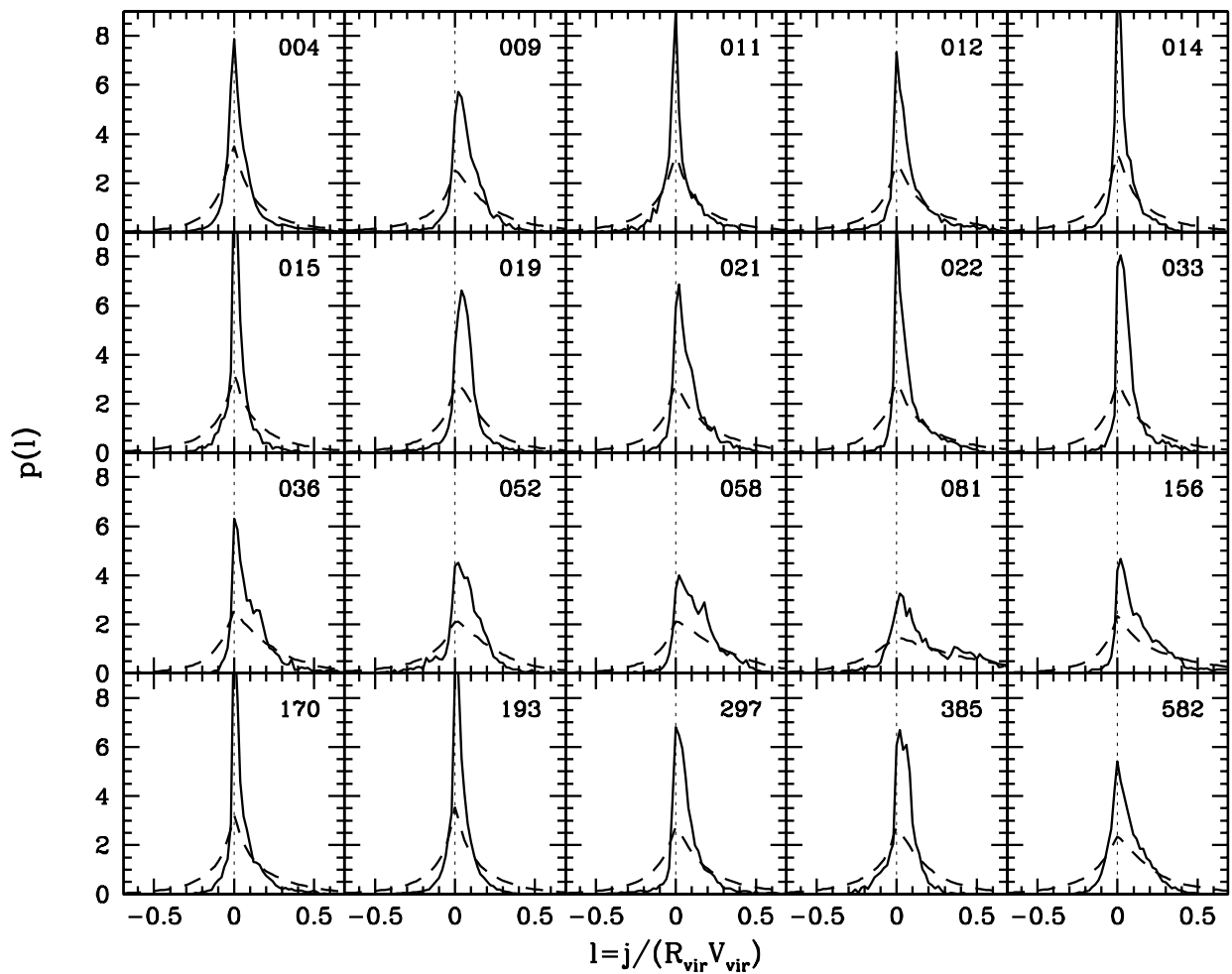}
\caption{A  comparison of  the angular  momentum  distributions $p(l)$
(solid lines) and  $p(l^v)$ (dashed lines) of the gas  in the 20 halos
listed in Table~1. Note how $p(l)$ typically has much smaller wings at
negative  specific angular  momentum  than $p(l^v)$  (cf, lower  right
panel of Figure~\ref{fig:lambda}).}
\label{fig:stream}
\end{figure}

In  Figure~\ref{fig:stream} we  compare the  gas AMD  $p(l^v)$ (dashed
lines) to the distribution $p(l)$ as computed from the {\it streaming}
motions of the  gas (solid lines). Here a  clear systematic difference
is apparent;  typically $p(l)$  has a less  extended wing  to negative
specific  angular momentum than  $p(l^v)$ (cf.   lower right  panel of
Figure~\ref{fig:lambda}), and  a more  pronounced peak at  low angular
momentum.  Note  that with equation~(\ref{spinpl})  both distributions
integrate to the same  value of $\lambda_{\rm gas}$. 

\begin{figure}
\plotone{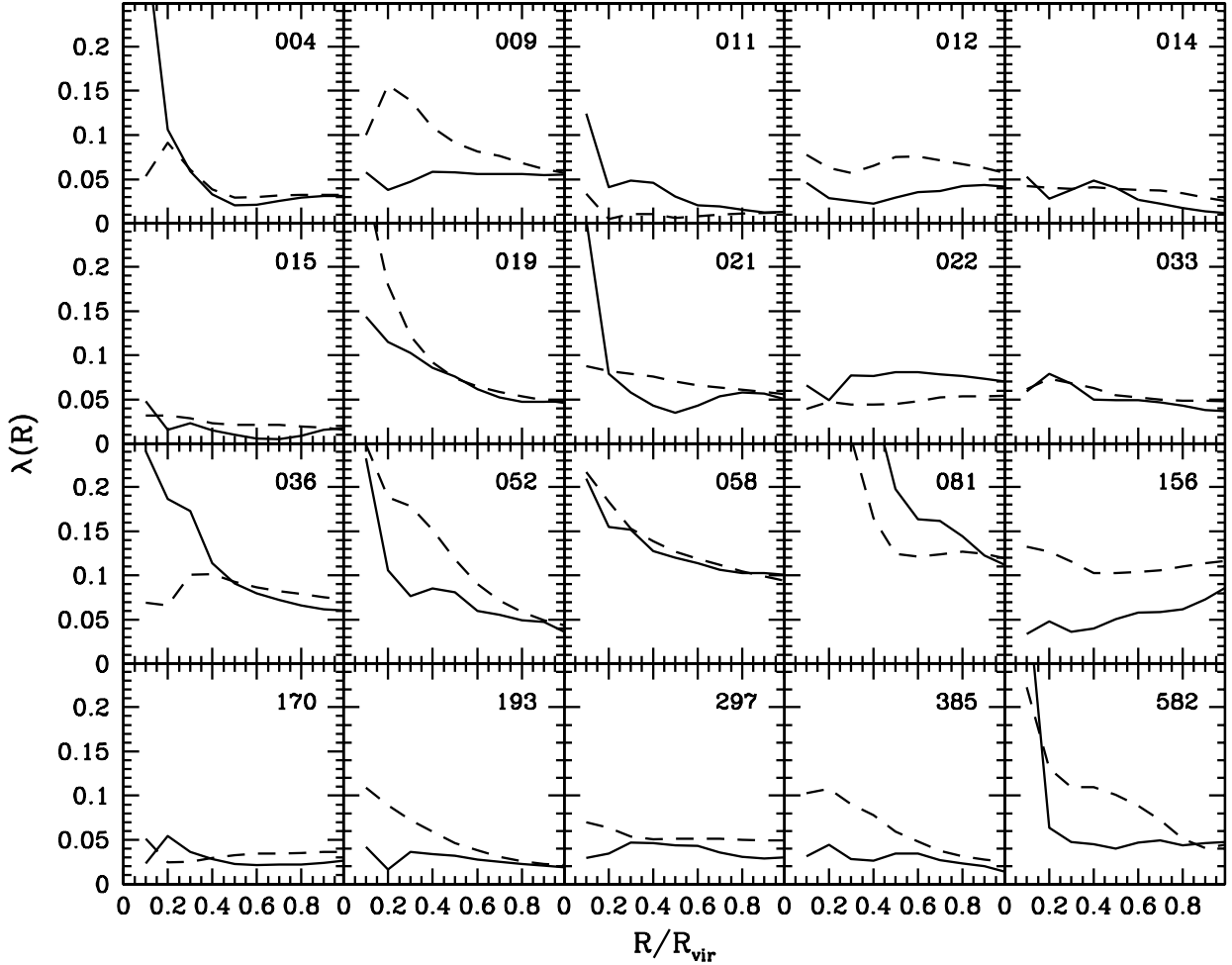}
\caption{The  values  of  the  spin  parameter inside  radius  $R$  as
functions of $R/R_{\rm vir}$ for the 20 halos listed in Table~1. Solid
(dashed)  lines  correspond  to  the  dark  matter  (gas).   Typically
$\lambda(R)$ either  decreases with radius or  stays roughly constant.
Although  in  some   halos  $\lambda_{\rm  DM}(R)$  and  $\lambda_{\rm
gas}(R)$  can  be  quite  different,  there is  no  indication  for  a
systematic  difference  between  $\lambda(R)$  of  the  gas  and  dark
matter.}
\label{fig:radprof}
\end{figure}

\begin{figure}
\plotone{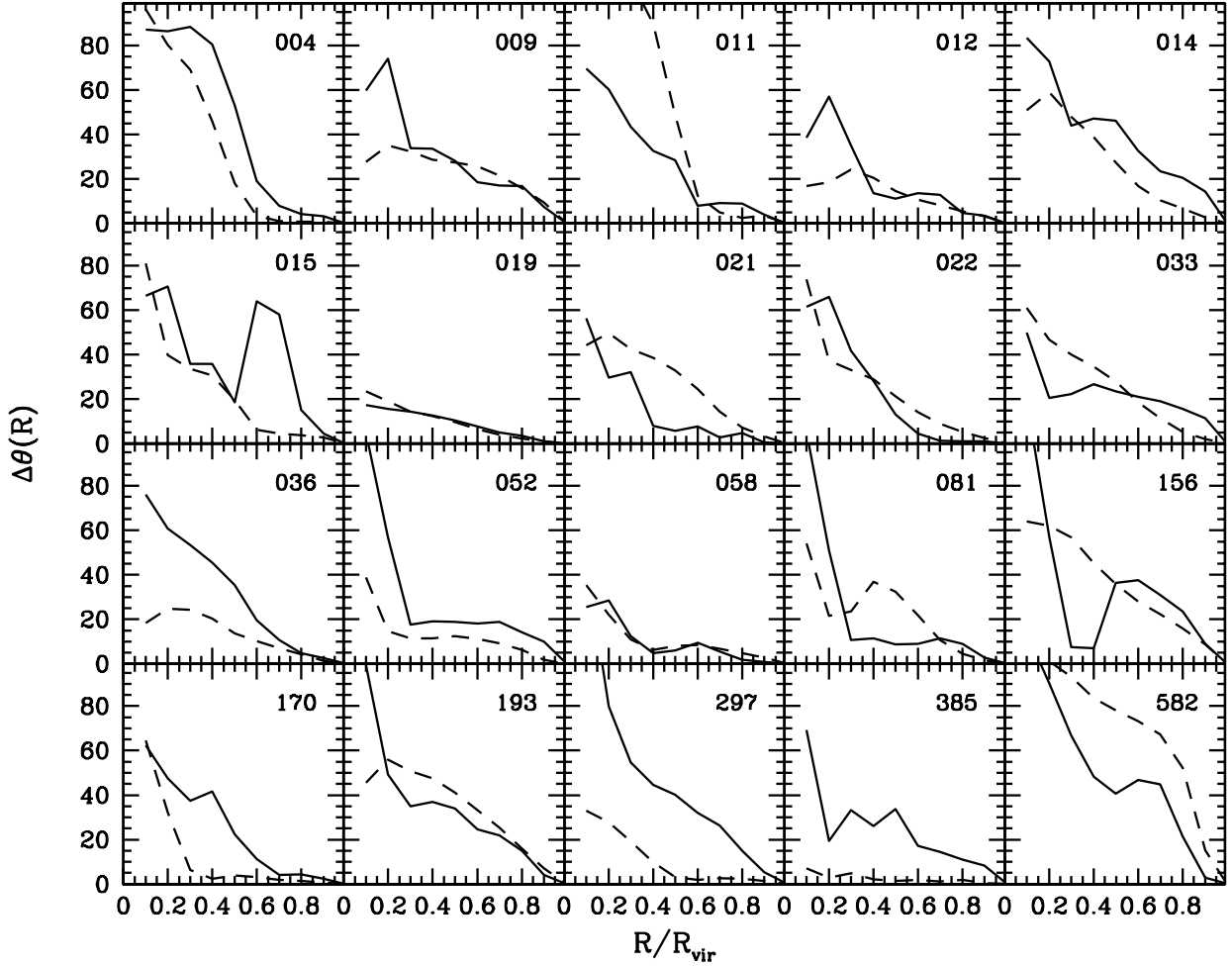}
\caption{The misalignment angle  $\Delta\theta(R)$ between the angular
momentum vector inside radius $R$ and  that of the entire halo.  As in
Figure~\ref{fig:radprof} solid  (dashed) lines correspond  to the dark
matter  (gas). Results  are plotted  as function  of  $R/R_{\rm vir}$.
Note that  the angular  momentum directions of  both the gas  and dark
matter can vary quite significantly  with radius, and in some cases is
clearly different for the gas and the dark matter.}
\label{fig:thetaprof}
\end{figure}

In    Figures~\ref{fig:radprof}   and~\ref{fig:thetaprof}    we   plot
$\lambda(R)$  and $\Delta\theta(R)$ as  functions of  $R/R_{\rm vir}$,
respectively.  Results for both the  dark matter (solid lines) and the
gas (dashed lines) are shown.  Here $\lambda(R)$ is the spin parameter
inside a sphere of radius $R$, i.e.,
\begin{equation}
\label{lamr}
\lambda(R) = {j^v(R) \over \sqrt{2 \, G \, R \, M(R)}}
\end{equation}
with $j^v(R)$ and $M(R)$ the specific angular momentum and mass inside
a sphere of radius  $R$, respectively. $\Delta\theta(R)$ is defined as
the angle  between the angular  momentum vector inside radius  $R$ and
that  inside $R_{\rm vir}$  (i.e., by  definition $\Delta\theta(R_{\rm
vir})=0$).  In some cases the  gas and dark matter have fairly similar
$\lambda(R)$ and  $\Delta\theta(R)$ profiles (e.g., halos  14, 19, and
58), while in  other cases the gas and dark matter  seem to have quite
distinct profiles  (e.g., halos 36, 156, 582).   Remarkably, halos for
which  the $\lambda(R)$  and/or $\Delta\theta(R)$  profiles  are quite
different for  the gas  and dark matter  are not necessarily  also the
ones that have the least similar AMDs, nor do they correspond to halos
for which  any of the global  parameters listed in Table~1  are in any
way special.  Note that  $\lambda$ typically either decreases or stays
roughly constant with radius, while the angular momentum directions of
both the gas  and the dark matter typically  change quite dramatically
from the center to the  virial radius.  We discuss the implications of
this  intrinsic misalignment  for the  formation of  disk  galaxies in
Section~\ref{sec:diskform}.

\subsection{Robustness of results}
\label{sec:robust}

The results  presented above are  based on a numerical simulation with
relatively low resolution ($128^3$  particles), with a small  box size
($L_{\rm box} = 10 h^{-1} \Mpc$ comoving), and which has only been run
to redshift $z=3$.  Too small  resolution  causes shocks to be  poorly
resolved, which could  result  in an   underestimate of the   possible
angular   momentum  redistribution of the    gas during its relaxation
process.  A small  box size may  cause an inaccurate representation of
the cosmological tidal field  responsible for the angular momentum  of
halos.  Finally,  it is   not  a priori  clear  that  angular momentum
distributions at $z=3$ are representative for those at $z=0$.

In  order to  address the   robustness  of our  results against  these
effects we analyze a high resolution numerical simulation of a cluster
kindly provided by Naoki Yoshida.  In this simulation exactly the same
cosmological parameters are  adopted  as in  the simulation  discussed
above  (see Section~\ref{sec:simul}). It is based  on the technique of
`zooming in' on a region of interest (Tormen,  Bouchet \& White 1997).
The   parent simulation from    which this region  is   chosen is  the
GIF-$\Lambda$CDM simulation of  Kauffmann \etal  (1999) which followed
$256^3$ particles within a comoving box with $L_{\rm box}=141.3 h^{-1}
\Mpc$.  The  Lagrangian region that collapses to  form the second most
massive halo in  this GIF simulation (with virial  mass $\sim 8 \times
10^{14}  h^{-1}  \Msun$) has  been  resimulated with greatly increased
mass and  force  resolution and including   a non-radiative gas.  This
high resolution region is represented by $\sim 2.2 \times 10^{5}$ dark
matter  and  gas particles each,  with masses  of $1.2  \times 10^{10}
h^{-1} \Msun$ and   $2.1  \times 10^{9} h^{-1}  \Msun$,  respectively.
Outside  this high-resolution region  $\sim 3.1 \times 10^6$ particles
with larger masses are used to represent the cosmological tidal field.
More details regarding this simulation  can be found in Springel \etal
(2001) and Yoshida \etal (2000a,b)

\begin{figure}
\plotone{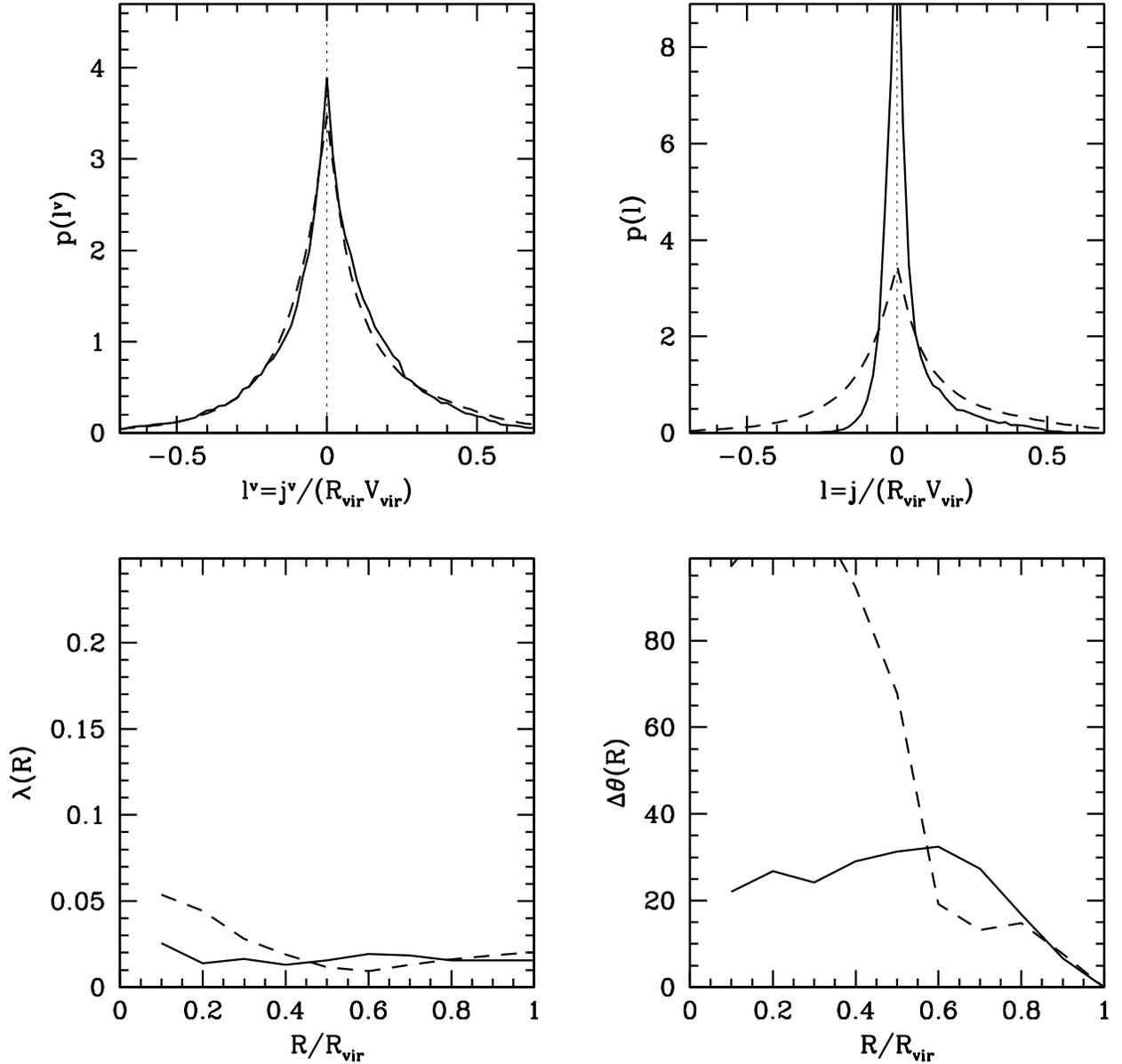}
\caption{Results  at  $z=0$  for  the high  resolution  simulation  of
cluster  `S0' described in  Section~\ref{sec:robust}.  The  upper left
panel plots $p(l^v)$ for both the dark matter (solid line) and the gas
(dashed line).  As for  the halos presented in Figure~\ref{fig:groups}
the dark matter and gas have virtually indistinguishable distributions
of specific  angular momentum. The  upper right panel  compares $p(l)$
(solid line) with $p(l^v)$  for the gas (cf. Figure~\ref{fig:stream}).
Note  that even  when considering  the streaming  motions of  the gas,
there is a  large fraction of gas mass  with negative specific angular
momentum (see  also Table~1). The  lower two panels  plot $\lambda(R)$
and $\Delta\theta(R)$  for halo `S0',  where the solid  (dashed) lines
correspond   to   the   dark   matter  (gas).    A   comparison   with
Figures~\ref{fig:radprof} and~\ref{fig:thetaprof} shows that halo `S0'
behaves      similar      as      the      halos      analyzed      in
Section~\ref{sec:distribution}.}
\label{fig:robust}
\end{figure}

We analyze the  angular momentum distributions of the  dark matter and
gas  of  this high  resolution  halo  at  $z=0$.  Using  the  standard
friends-of-friends algorithm  (Davis \etal 1985) we  identify the most
massive group  and its  most bound particle.   Next the  virial radius
$R_{\rm vir}$ of  the spherical volume, centered on  this particle, is
determined  adopting $\Delta_{\rm  vir} =  101$ (appropriate  for this
cosmology and redshift; Bryan \&  Norman 1998).  The halo thus defined
(hereafter referred to  as halo `S0') contains $\sim  1.2 \times 10^5$
particles (more  than twice that of  the biggest halo  analyzed in the
previous section), and  has a virial mass and  radius of $M_{\rm vir}=
8.7 \times 10^{14} h^{-1} \Msun$ and $1.86 h^{-1} \Mpc$, respectively.
Analyzing the  angular momentum distributions  of the dark  matter and
gas  particles  as  in  Section~\ref{sec:angmom} we  obtain  the  spin
parameters and mass fractions  with negative specific angular momentum
listed  in  Table~1.   The  resulting angular  momentum  distributions
$p(l^{v})$ and $p(l)$, as  well as $\lambda(R)$ and $\Delta\theta(R)$,
are shown in Figure~\ref{fig:robust}.

Note that  the angular momentum distributions of  halo  `S0' are in no
way  systematically  different  from those  of    the sample of  halos
analyzed above.  In particular, once  again we  find  that the gas and
dark  matter have virtually  identical angular  momentum distributions
$p(l^v)$, and  that a large  mass fraction ($\sim  45$ percent) of the
gas  has  negative specific  angular  momentum.   The $\lambda(R)$ and
$\Delta\theta(R)$ profiles  are    similar  to   that  of halo      11
(cf. Figures~\ref{fig:radprof} and~\ref{fig:thetaprof}).  Although one
has to   be  careful  to draw   conclusions based  on  a   single high
resolution halo, these results suggest that the characteristics of the
angular        momentum      distributions        presented         in
Section~\ref{sec:distribution}  are not  influenced by the  relatively
low resolution, by the small box size, or by the fact that we analyzed
the halos at $z=3$.

\subsection{Do Dark Matter Halos follow a Universal Angular Momentum
Profile?}
\label{sec:universal}

Bullock  \etal (2001;  hereafter  B01) computed  the angular  momentum
distributions  of dark  matter halos  and argued  that these  follow a
universal profile given by:
\begin{equation}
\label{mb00}
P(j) \equiv  {M(<j) \over  M_{\rm vir}} =  
{\mu (j/j_{\rm  max}) \over (j/j_{\rm max}) + \mu - 1},
\end{equation}
Here $M(<j)$ is the halo mass with specific angular momentum less than
$j$, $j_{\rm  max}$ is  the maximum specific  angular momentum  of the
entire AMD, and  $\mu$ is a free fitting  parameter.  Note that $P(j)$
corresponds  to  the {\it  cumulative}  distribution  of the  specific
angular momentum.  In order to  compute these AMDs, B01 split up their
halos in  a number  of spatial cells.   The angular  momentum profiles
$P(j)$ are  then constructed by  ranking the cells according  to their
total specific angular momentum  $j$ from which the best-fitting value
of  $\mu$ is  determined.  Note  that  this cell  averaging should  in
principle  average out  the random  motions  $\w$ of  the dark  matter
particles, and equation~(\ref{mb00}) should thus be interpreted as the
cumulative distribution of $p(j)$ rather than $p(j^v)$.

However, there are two problems. First of all, it is unlikely that the
particular  cell geometry  adopted by  B01 yields  the  true streaming
motions    of   the    dark    matter   (see    the   discussion    in
Section~\ref{sec:theory}).  This  means that, depending  on the actual
cell  geometry  adopted, the  AMD  of  equation~(\ref{mb00}) does  not
necessarily   reflect  the   true  $p(j)$.    Secondly,   because  $j$
corresponds to a projected component the  total $j$ of any cell can be
negative. B01  ignored this and removed all   cells with negative $j$
from their ranked list. We find that the $p(l)$  of our halos are well
fit  by  equation~(\ref{mb00}) for  $l   \geq  0$, in  agreement  with
B01. However, whereas  B01 have implicitely assumed  that there is  no
mass with negative   specific  angular momentum,  we  find large  mass
fractions with $l < 0$, at least for the gas. Since measuring the true
streaming  motions  of  the dark matter  is  non-trivial  we  have not
attempted  to obtain the $p(l)$ for   the dark matter.  However, since
the $p(l^v)$ of  the gas and dark  matter are virtually  identical, it
seems   reasonable  to   assume   that  the  same   applies    for the
$p(l)$-distributions\footnote{The accuracy of this assumption needs to
be  checked using   simulations  of higher   resolution  than the  one
analyzed here.}.  If confirmed, between $\sim 5$ and 50 percent of the
dark matter mass   has negative specific  angular  momentum,  in clear
disagreement with the `Universal'  profile of B01.  Furthermore, since
the mass fraction with negative specific  angular momentum is strongly
anti--correlated  with the    total  specific angular  momentum,   the
`Universal' angular momentum profile of equation~(\ref{mb00}) lacks an
essential characteristic of the true AMDs.
 
\section{Implications of the Formation of Disk Galaxies}
\label{sec:diskform}

In the  standard picture of  disk formation, (i) the  specific angular
momentum of  disk galaxies arises  from cosmological torques,  (ii) is
identical to that  of their dark matter halos,  and (iii) is conserved
during the  process of disk formation.  Recently  several authors have
pointed out that this picture is inconsistent with observations, as it
would result  in disks that  are too centrally concentrated  (B01; van
den Bosch 2001; van den Bosch \etal 2001). This is a reflection of the
fact that  the specific angular momentum distributions  of dark matter
halos,  as  described  by  the  `Universal' profile  of  B01,  contain
relatively too much low angular momentum material.

As we argued in Section~\ref{sec:intro} this seems to suggest that one
or more of the standard  assumptions is incorrect. Our results clearly
show that in the non-radiative case the angular momentum distributions
of the  gas and dark  matter are virtually identical,  thereby lending
strong  support to  (ii).   However, our  results  also indicate  that
assumption  (iii) cannot be  correct. As  shown, the  angular momentum
distributions  $p(l)$  of  the   gas  contain  relatively  large  mass
fractions with negative specific  angular momentum.  Disk galaxies, on
the  other  hand, generally  do  not  contain  material with  negative
specific angular  momentum.  This suggests  that detailed conservation
of specific angular momentum has to  be violated if one wants to build
a disk out of the typical AMDs of the gas in proto-galaxies.

Once  the gas  starts to  cool, its  density increases,  enhancing the
frequency  of shocks through  which the  gas parcels  exchange angular
momentum.   Gas parcels with  negative $j_z$  can share  their angular
momentum with  parcels with positive  $j_z$, thus increasing  the mass
fraction with low angular momentum.  This, however, can only aggravate
the  low  angular momentum  problem.   Apparently  real  disks do  not
contain a  large fraction  of low angular  momentum material  (van den
Bosch \etal  2001).  It is  tempting, however, to speculate  that this
low  angular  momentum  material  might actually  build  a  (near-zero
angular  momentum) bulge component  (cf.  van  den Bosch  1998; Kepner
1999).  The mass of such a bulge depends on the details of the angular
momentum  exchange during  cooling.  The  mass with  negative specific
angular  momentum may  combine with  very  little mass  but with  high
positive specific angular momentum (yielding a minimum bulge mass), or
with  a relatively  large amount  of  mass with  small, but  positive,
specific  angular momentum  (yielding  a maximum  bulge  mass).  As  a
simple  intermediate case,  consider the  scenario in  which  the zero
angular momentum bulge forms out of the material with AMD
\begin{equation}
\label{plbulge}
p_{\rm bulge}(l) = \left\{ \begin{array}{ll}
                            p(l)  & l < 0 \\
                            p(-l) & l \geq 0 
                            \end{array} \right.
\end{equation}
If the  remaining material builds a disk,  the resulting bulge-to-disk
mass ratio is
\begin{equation}
\label{btod}
B/D = {2  f_{\rm gas} \over 1  - 2 f_{\rm  gas}}, 
\end{equation}
while the angular momentum distribution  of the disk material is given
by $p_{\rm disk}(l) \equiv p(l) -  p(-l)$ where $l \geq 0$.  Under the
standard assumption that this  material conserves its specific angular
momentum,  and that  the dark  matter follows  an NFW  density profile
(Navarro, Frenk \& White 1997), the surface density $\Sigma(R)$ of the
resulting disk component can be computed using the relation
\begin{equation}
\label{diskform}
{M_{\rm disk} \over 1-2 f_{\rm gas}}
\int_{l_1}^{l_2} p_{\rm disk}(l) \, {\rm d}l = 
2 \, \pi \int_{R_1}^{R_2} \Sigma(R) \, R \, {\rm d}R
\end{equation}
where $l$ and $R$ are related according to
\begin{equation}
\label{dform}
l = {R \, V_c(R) \over R_{\rm vir} \, V_{\rm vir}} = 
\sqrt{ x {{\rm ln}(1+c x) - c x / (1 + c x) \over 
{\rm ln}(1+c) - c / (1 + c)}}
\end{equation}
with $x=R/R_{\rm vir}$, $V_c(R)$  the circular velocity at radius $R$,
and  $c$ the  NFW concentration  parameter for  which we  adopt $c=10$
throughout.  For simplicity we ignore the self-gravity of the disk, so
that $V_c(R)$ reflects the circular velocity of the NFW halo.

\begin{figure}
\plotone{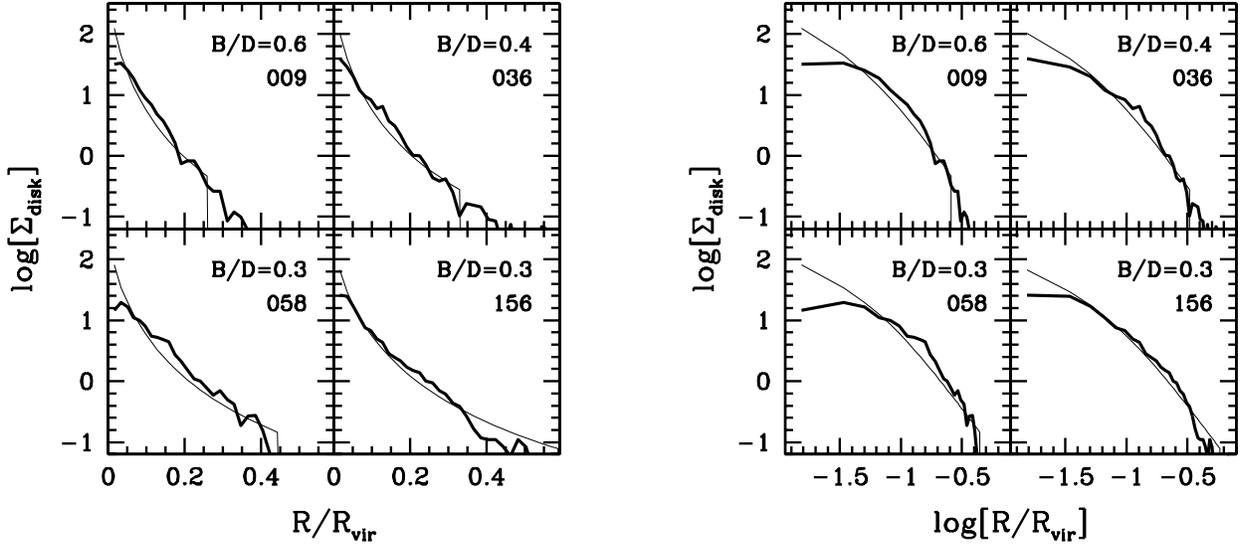}
\caption{The thick solid lines correspond to the density distributions
(in arbitrary  units) of the disks  that form in four  halos listed in
Table~1 under the assumptions that (i) material with negative specific
angular  momentum forms  a zero-angular  momentum bulge  together with
material  with  the  same,  but positive,  specific  angular  momentum
distribution  (e.g.,  equation~[\ref{plbulge}]),  (ii)  the  remaining
material forms  a disk  under detailed angular  momentum conservation,
(iii)  the dark  matter  halo  has an  NFW  density distribution  with
$c=10$, and  (iv) the  self-gravity of the  disk can be  ignored.  The
resulting  bulge-to-disk  mass  ratios  $B/D$ are  indicated  in  each
panel. The  thin solid lines correspond  to disks that form  out of an
angular  momentum distribution  given  by the  `Universal' profile  of
equation~(\ref{buldiff})  with $\mu=1.25$  and $\lambda_{\rm  gas}$ as
listed in Table~1. We plot  $\Sigma_{\rm disk}(R)$ both as function of
$R/R_{\rm vir}$  (panels on the  left) and ${\rm  log}(R/R_{\rm vir})$
(panels on the right), in  order to emphasize the differences at large
and small radii, respectively.}
\label{fig:disks} 
\end{figure}

In Figure~\ref{fig:disks}  we plot the resulting  disk surface density
distributions  $\Sigma(R)$  for  four  representative halos  from  our
sample  of 23  (thick solid  lines).  In  each panel  we  indicate the
bulge-to-disk  ratio computed  with  equation~(\ref{btod}).  The  thin
lines correspond to  disks that would form out  of an angular momentum
distribution given by
\begin{equation}
\label{buldiff}
p(l) = {\zeta \mu (\mu - 1) \over (\zeta l + \mu - 1)^2},
\end{equation}
which corresponds to the `Universal' profile of equation~\ref{mb00}.
Here 
\begin{equation}
\label{zeta}
\zeta = {1 \over \sqrt{2} \lambda_{\rm gas}} \left( 1 - \mu \left[ 1 - 
(\mu - 1) {\rm ln} \left( {\mu \over \mu - 1}\right) \right] \right).
\end{equation}
where $\lambda_{\rm gas}$ is the spin parameter of the gas in the halo
under consideration  (listed in  Table~1). We adopt  $\mu=1.25$, which
corresponds to the median value of all halos analyzed by B01.

Whereas the  disks forming out  of the AMDs  of the gas  analyzed here
have surface  brightness profiles that  are very close  to exponential
(corresponding to a straight line in  the panels on the left), this is
not  the case  for  disks forming  out  of the  `Universal' AMD.   The
largest  difference between  the  two surface  brightness profiles  is
apparent at small radii.  This  is due to the bulge-formation scenario
included in one case  but not the other. In fact, as  shown by van den
Bosch (2001),  if a simple  bulge formation scenario is  included, the
`Universal' AMD also yields  near-exponential disks.  One problem that
is not solved  by turning low angular momentum  material into a bulge,
however,  is the fact  that the  `Universal' AMD  yields disks  with a
clear truncation radius which occurs  at too high surface density (van
den Bosch 2001). However, the disks forming out of the AMDs of the gas
analyzed  here   continue  as  exponentials  to   much  lower  surface
brightness,  suggesting  that  the   cell  averaging  adopted  by  B01
underestimated the true $j_{\rm max}$.

Thus, the angular momentum  distributions of the gas in proto-galaxies
seem  consistent with  the observed  surface density  distributions of
disk galaxies if the  material with negative specific angular momentum
is assumed  to make a bulge component.  Nevertheless, several problems
remain.

First of all,  since $0.05 \lta f_{\rm gas} \lta  0.5$ in the scenario
considered  above, a  large  fraction  of the  halos  will form  bulge
dominated  `early-type' systems. If  we use  a bulge-to-disk  ratio of
$B/D=1$  to discriminate  between late  and early-type  galaxies, disk
dominated late  type galaxies  only form in  $\sim 60$ percent  of all
halos in  our sample.   Furthermore, with a  minimum $f_{\rm  gas}$ of
about 5  percent (see  Figure~\ref{fig:lambda}), no systems  will form
with $B/D  < 0.1$,  so the  question of how  bulge-less dwarf  and LSB
galaxies  form (van  den  Bosch 2001;  van  den Bosch  \etal 2001)  is
unanswered.

Another  problem   concerns  the  various   misalignments  of  angular
momentum.    As    we   have   shown   in   the    lower   panels   of
Figure~\ref{fig:radprof},  not  all  gas  has the  same  direction  of
angular momentum.  When the self-gravity of the gas becomes important,
gas  parcels with  different  angular momentum  directions (moving  on
non-coplanar  circular  orbits)  will  exert torques  on  each  other,
resulting in a redistribution  of angular momentum.  In addition, when
the  gas  is  misaligned  with   the  potential  of  the  dark  matter
halo\footnote{We emphasize here that any possible misalignment between
the  gas and the  moment of  inertia of  the dark  matter halo  is not
related  to  the  misalignment  angle  $\theta$  between  the  angular
momentum  vectors  of   the  gas  and  dark  matter:   halos  are  not
rotationally  supported systems, and  as shown  by Porciani,  Dekel \&
Hoffman (2002), the  spin and inertia of dark  matter halos are poorly
aligned.},  transfer of  angular  momentum between  the  gas and  dark
matter will occur. If the gas predominantly transfers angular momentum
to the dark matter, which one might naively expect since the latter is
the dominant mass component, this  results in an additional problem as
many authors  in the past  have pointed out:  In order to  explain the
observed distribution of  disk scale lengths the gas  cannot lose much
of its angular momentum.

\section{Conclusions}
\label{sec:concl}

It  is  generally assumed  that  the  gas and  the  dark  matter in  a
proto-galaxy have the same  distribution of specific angular momentum.
This is motivated by the fact that both mass components experience the
same  tidal  torques.   However,  the two  components  undergo  rather
different  relaxation mechanisms:  whereas the  dark  matter undergoes
collisionless, violent relaxation, the gas gets shocked which could in
principle cause a redistribution of its angular momentum distribution.
In  this paper we  have used  numerical simulations  to test,  for the
first time,  whether the  {\it ansatz} that  gas and dark  matter have
similar  initial  angular   momentum  distributions  is  correct.   We
presented  a   numerical  simulation  of  structure   formation  in  a
$\Lambda$CDM cosmology including both  dark matter and a non-radiative
gas.   For $~\sim  380$ halos  with virial  masses in  the  range $4.8
\times  10^9 h^{-1}  \Msun \leq  M_{\rm vir}  \leq 1.1  \times 10^{12}
h^{-1} \Msun$ (at  $z=3$) we computed the angular  momenta of both the
gas and the dark matter.

We have  shown that on average the  gas and dark matter  have the same
distribution  of  spin  parameters  and that  their  detailed  angular
momentum  distributions in  individual halos  are  remarkably similar.
Evidently, shocking during the virialization process does not decouple
the  angular  momentum  of the  gas  from  that  of the  dark  matter,
supporting  the standard  assumption  that gas  and  dark matter  have
identical  angular momentum  distributions. We  have shown  that these
results  are robust  by  also analyzing  a  high resolution  numerical
simulation of  a single cluster-sized halo  at $z=0$ in  a much larger
cosmological volume.

A  careful analysis  of  the detailed  angular momentum  distributions
reveals  that  they  are  inconsistent with  the  `Universal'  profile
suggested  by B01.   In the  angular momentum  distributions presented
here  between 5  and  50 percent  of  the mass  has negative  specific
angular momentum, whereas B01  have (implicitly) assumed that all mass
has  positive  specific angular  momentum.   Furthermore, our  angular
momentum distributions  have wings out  to very high  specific angular
momentum,  whereas the profile  suggested by  B01 has  a cut-off  at a
certain  $j_{\rm max}$  that results  in disk  truncation at  too high
surface density (van den Bosch 2001).

The fact  that relatively large mass fractions  have negative specific
angular  momentum suggests  that the  standard assumption  of detailed
specific  angular  momentum  conservation,  which  is  often  made  in
theories  of  disk formation,  cannot  be  correct.   After all,  disk
galaxies  do not  typically  contain material  with negative  specific
angular momentum.   We have suggested that during  the cooling process
the gas with negative specific angular momentum may collide/shock with
material  with positive  specific angular  momentum to  build  a bulge
component with zero  angular momentum. In such a  picture at least the
{\it total} amount of angular momentum of the gas is conserved, and we
have  shown that it  can produce  disks with  near-exponential surface
density distributions  without the  small truncation radii  implied by
the  `Universal'  profile  of  B01.   Numerical  simulations  of  disk
formation that include cooling do  indeed seem to reveal the formation
of a dense central knot of gas  in addition to the disk (e.g., Katz \&
Gunn 1991; Navarro \& White 1994).  Although this `bulge' is typically
interpreted  as forming  out of  sub-clumps that  lose  their specific
angular momentum  to the dark  matter through dynamical  friction, our
results  suggest  that even  without  substructure  a  bulge may  form
naturally.  In  fact, taking the  results at face value  suggests that
$\sim  40$ percent  of all  halos are  unable to  form  disk dominated
galaxies,  simply because  they  have too  large  mass fractions  with
negative  specific  angular   momentum.   In  addition,  systems  with
bulge-to-disk mass ratios below 10 percent will be extremely rare.

In reality the formation of  disk galaxies will be more complicated as
depicted above.   We have shown  that the angular momentum  vectors of
the gas  and dark  matter are misaligned  by on average  $\sim 35^{\rm
o}$. Surprisingly the amount of  misalignment is not a measure for the
degree to which the respective angular momentum distributions agree or
disagree  with  each  other.   There   is  a  weak  tendency  for  the
misalignment  to be  weaker in  halos that  are more  massive  or have
larger spin parameters. Higher  resolution simulations are required to
investigate the amount and origin of this misalignment in more detail.
In  addition to  this  misalignment between  the  two components,  the
angular  momentum  vector  of  each individual  component  can  change
direction   quite  dramatically  with   radius.   This   is  typically
interpreted as  originating from  decoherence in the  angular momentum
orientation of material accreted at different times (Ryden 1988; Quinn
\& Binney 1992) and is often invoked as a mechanism to create warps in
disk  galaxies (e.g., Ostriker  \& Binney  1989).  Although  the exact
implications of these various misalignments are currently unclear, and
require  a   more  detailed  investigation,  it  is   clear  that  any
misalignment of the gas with  either itself or with the inertia moment
of the dark matter  halo has to vanish if one wants  to build a stable
disk.  If this  results in a significant transfer  of angular momentum
from  the gas  to  the dark  matter  these misalignments  might be  an
additional cause  for the  angular momentum catastrophe  hampering the
formation  of   sufficiently  extended  disk   galaxies  in  numerical
simulations.

Finally, we  emphasize that the implications  for disk/bulge formation
discussed above  are highly speculative.   In our simulations  the gas
was  not   allowed  to  cool,   and  we  used  the   angular  momentum
distributions of the gas found at $z=3$ to speculate about the fate of
the gas once cooling were turned  on.  In reality the gas will already
have  (partially) cooled  inside  sub-clumps that  merge  to form  the
systems  analyzed here,  and  this  will also  impact  on the  angular
momentum distribution  of the gas and its  subsequent evolution (e.g.,
dynamical friction).  On the other hand, it has been argued by several
authors that in order to  prevent the angular momentum catastrophe the
gas has to be prevented from cooling inside sub-clumps for instance by
feedback processes  or preheating (e.g., Navarro \&  White 1994; Weil,
Eke \&  Efstathiou 1998; Sommer-Larsen, Gelato \&  Vedel 1999).  Under
those circumstances  the angular momentum distribution of  the gas may
in fact resemble  those analyzed here, although it  is unclear to what
extent  these   heating  processes  influence   the  angular  momentum
distributions of  the gas  prior to being  incorporated in  the halos.
For instance,  processes related to feedback  and re-ionization create
pressure gradients in the gas  which may decouple the angular momentum
distribution of the  gas from that of the dark  matter. In a follow-up
paper  (Abel \etal 2002)  we address  in more  detail the  impact such
pressure  forces may  have on  the angular  momentum  distributions in
proto-galaxies.

\acknowledgments

We  are grateful  to Naoki  Yoshida for providing  his high resolution
numerical simulation,    to James  Bullock,  Andreas  Burkert, Avishai
Dekel, Ariyeh Maller, Joel  Primack and Risa Wechsler  for stimulating
discussions, and to Volker Springel for his help with the simulations.
This work has partially been  supported through NSF grants ACI96-19019
and AST-9803137.  We also acknowledge support from the Grand Challenge
Cosmology Consortium, and from the EC RTN network ``The Physics of the
Intergalactic Medium''.

\clearpage

\clearpage

\begin{table}
\begin{center}
\caption{Global Parameters of Halos in
 Figures~\ref{fig:groups}--\ref{fig:thetaprof}.\label{tab:haloes}}
\begin{tabular}{crrcccccr}
\tableline\tableline
ID & $N_{\rm DM}$ & $N_{\rm gas}$ & $\lambda_{\rm DM}$ & $\lambda_{\rm
gas}$ & $f^{v}_{\rm DM}$ & $f^{v}_{\rm gas}$ & $f_{\rm gas}$ & $\theta$ \\
(1) & (2) & (3) & (4) & (5) & (6) & (7) & (8) & (9) \\
\tableline
$004$ & $28584$ & $27099$ & $0.031$ & $0.032$ & $0.458$ & $0.445$ & $0.386$ &  $3.5$ \\
$009$ &  $3137$ &  $3196$ & $0.055$ & $0.057$ & $0.378$ & $0.375$ & $0.186$ & $45.7$ \\
$011$ &  $4951$ &  $4551$ & $0.013$ & $0.012$ & $0.438$ & $0.475$ & $0.471$ & $57.2$ \\
$012$ &  $3989$ &  $3986$ & $0.041$ & $0.057$ & $0.440$ & $0.389$ & $0.239$ &  $0.9$ \\
$014$ &  $7394$ &  $6804$ & $0.012$ & $0.025$ & $0.477$ & $0.442$ & $0.312$ & $45.8$ \\
$015$ &  $3194$ &  $2901$ & $0.018$ & $0.016$ & $0.449$ & $0.458$ & $0.313$ &  $4.4$ \\
$019$ &  $8555$ &  $8111$ & $0.048$ & $0.045$ & $0.363$ & $0.367$ & $0.171$ &  $6.0$ \\
$021$ &  $3878$ &  $3726$ & $0.050$ & $0.056$ & $0.390$ & $0.380$ & $0.186$ & $15.9$ \\
$022$ &  $2882$ &  $2675$ & $0.070$ & $0.054$ & $0.363$ & $0.400$ & $0.204$ & $20.3$ \\
$033$ &  $2600$ &  $2656$ & $0.037$ & $0.048$ & $0.392$ & $0.402$ & $0.140$ & $25.3$ \\
$036$ &  $2786$ &  $2582$ & $0.060$ & $0.073$ & $0.345$ & $0.341$ & $0.147$ &  $2.8$ \\
$052$ &  $2684$ &  $2527$ & $0.035$ & $0.043$ & $0.389$ & $0.389$ & $0.245$ & $29.5$ \\
$058$ &  $3652$ &  $3624$ & $0.100$ & $0.094$ & $0.272$ & $0.299$ & $0.118$ &  $5.3$ \\
$081$ &  $2726$ &  $2653$ & $0.111$ & $0.119$ & $0.317$ & $0.333$ & $0.209$ &  $9.6$ \\
$156$ &  $4011$ &  $3825$ & $0.087$ & $0.117$ & $0.373$ & $0.312$ & $0.112$ &  $9.5$ \\
$170$ &  $3660$ &  $3534$ & $0.027$ & $0.036$ & $0.415$ & $0.429$ & $0.269$ &  $8.5$ \\
$193$ & $21030$ & $20208$ & $0.019$ & $0.021$ & $0.444$ & $0.444$ & $0.283$ & $46.8$ \\
$297$ &  $9582$ &  $9218$ & $0.030$ & $0.049$ & $0.437$ & $0.401$ & $0.222$ &  $5.1$ \\
$385$ &  $3187$ &  $2962$ & $0.013$ & $0.025$ & $0.457$ & $0.422$ & $0.246$ & $16.8$ \\
$582$ & $17321$ & $16719$ & $0.048$ & $0.044$ & $0.414$ & $0.411$ & $0.310$ & $30.4$ \\
 S0   & $63171$ & $56487$ & $0.016$ & $0.022$ & $0.461$ & $0.474$ & $0.451$ & $61.5$ \\
\tableline
\end{tabular}

\tablecomments{Global parameters  of the halos  whose angular momentum
distributions               are              presented              in
Figures~\ref{fig:groups}--\ref{fig:thetaprof}.   Column (1)  lists the
halo ID.  Columns~(2)  and~(3) list the number of  dark matter and gas
particles,  Columns~(4) and~(5) the  spin parameters,  and Columns~(6)
and~(7)  the mass  fractions  with negative  $j^v_z$. For  comparison,
column~(8) lists the gas mass fraction with negative $j_z$ .  Finally,
column~(9) lists the misalignment angle $\theta$.}

\end{center}
\end{table}

\end{document}